\documentclass[12pt]{article}
\usepackage{epsfig}

\begin{document}

\hskip 12 cm 
{\Large RUB-TP2-19/03}

\begin{center}
{\Large {\bf DVCS on spinless nuclear targets in impulse approximation}}
\vskip 0.5cm 
V. GUZEY\\
{\it Institut f{\"u}r Theoretische Physik II, Ruhr-Universit{\"a}t Bochum,
  D-44780 Bochum, Germany}\\
\vskip 0.25cm
M. STRIKMAN\\
{\it Department of Physics, the Pennsylvania State University, State
  College, PA 16802, USA}
\end{center}

{\abstract Within the impulse approximation, we derive expressions for
 the amplitude of deeply virtual Compton scattering on spinless
 nuclei   in terms of the generalized parton distributions of the
 nucleon. As an application, nuclear
  effects in the beam-charge and single-spin asymmetries are  discussed.}

\section{Introduction}

Deeply Virtual Compton Scattering (DVCS) attracts a large amount of
interest, both experimentally and theoretically. The initial results on
DVCS have been reported by HERMES \cite{HERMES}, ZEUS \cite{ZEUS} and
H1 \cite{H1} experiments at DESY and
the CLAS experiment \cite{CLAS} at TJNAF; the recent progress in the theoretical
understanding of DVCS is summarized in the review articles \cite{BMK02},\cite{GPV01},\cite{GV98}.   

DVCS on a given target accesses new non-perturbative objects --
Generalized Parton Distributions (GPDs) of the target. These
distributions contain encoded information on the quark-gluon structure
of the target.

 To date most of the
experimental and theoretical work dedicated to DVCS concern with
nucleon and pion targets. However, there are already first data on DVCS
on nuclear (Deuterium, Neon) targets taken at HERMES  \cite{Ellinghaus}. On the
theoretical side, nuclear GPDs may allow one to better understand the
nature of nuclear forces \cite{Polyakov}.

Previously, the impulse approximation approach to 
nuclear effects in  coherent
DVCS on Deuterium and other
spin-1 nuclei was considered in Refs.~\cite{Cano}.
The aim of this paper is to apply the impulse approximation to DVCS on
spin-0 nuclei and, in particular, to obtain a qualitative estimate of
the Fermi motion effect on
the beam-charge asymmetry for DVCS a spinless nucleus of Neon that was
recently measured at HERMES \cite{Ellinghaus}.

In order to make clear the principal steps of the derivation,
we first consider the calculation of the electromagnetic form factor of
a spinless nucleus.

\section{Electromagnetic form factor of a spinless nucleus}
\label{sec:em}

The electromagnetic (charge) form factor of a spinless nucleus,
$F^{e.m.}_{A}(t)$, is defined using the matrix element of the
operator of the electromagnetic current, $J^{\mu}(0)$, between the states
of the initial and final nucleus with momenta $P_{A}$ and
$P_{A}^{\prime}=P_{A}+q$ ($t \equiv q^2 < 0$):
\begin{equation}
\langle P_{A}+q
|J^{\mu}(0)|P_{A}\rangle=(2P_{A}+q)^{\mu}F^{e.m.}_{A}(t) \, .
\label{eq:def}
\end{equation} 
The matrix element in Eq.~(\ref{eq:def}) can be evaluated using the
impulse approximation which is based on the intuitive picture that the
interaction with the nucleus is a two-step process: First, the nucleus
is decomposed into non-relativistically moving nucleons; then, these
nucleons interact independently  with the probe. 
As a result, the matrix elements between the nuclear states can be
expressed in terms of the nucleon matrix elements.

The consistent derivation of the impulse approximation 
and discussion of its limitations was given
in Ref.~\cite{FS81}.
 In what follows, we shall present a less rigorous treatment of
the impulse approximation.
It is convenient to
start from the covariant expression for the matrix element in
Eq.~(\ref{eq:def}) (see Ref.~\cite{Sargsian})
\begin{eqnarray}
&&\langle P_{A}+q|J^{\mu}(0)|P_{A}\rangle=\sum_{{\rm nucleons}}\int
\frac{d^4 p_1}{i (2\pi)^4} \dots \frac{d^4 p_A}{i (2\pi)^4}
\bar{\Gamma}_{A}(p_1+q, p_2, \dots,p_A) \nonumber\\
&&\times 
\Big(\frac{-1}{\hat{p_1}+\hat{q}-m}\Big)\hat{\Gamma}_{e.m.}^{\mu}(q)\Big(\frac{-1}{\hat{p_1}-m}\Big)
\frac{(-1)^{A-1}}{(\hat{p_2}-m) \dots (\hat{p_A}-m)}\Gamma_{A}(p_1,
  p_2, \dots, p_A) \nonumber\\
&&\times i (2\pi)^4 \delta^4(P_A-p_1-\dots -p_A) \,,
\label{eq:long}
\end{eqnarray} 
with $\sum_{{\rm nucleons}}$ the incoherent sum of
electromagnetic interactions with each nucleon,
$\bar{\Gamma}_{A}$ and $\Gamma_{A}$ the covariant vertex functions
describing the transition of $A$ nucleons to the nucleus, and
$\hat{\Gamma}_{e.m.}^{\mu}$ the electromagnetic vertex of the nucleon,
\begin{equation}
\hat{\Gamma}_{e.m.}^{\mu}(q)=\gamma^{\mu}F_1(t)+\frac{i \sigma^{\mu \nu}
  q_{\nu}}{2 m} F_2(t)  \,,
\label{eq:gamma}
\end{equation}
where $F_1$ and $F_2$ are the nucleon elastic form factors; $m$ is the
nucleon mass.

The covariant vertices $\bar{\Gamma}_{A}$ and $\Gamma_{A}$ in
Eq.~(\ref{eq:long}) can be related to  non-relativistic nuclear wave
function of the target.
This  non-relativistic 
 reduction of Eq.~(\ref{eq:long}) can be carried out
as follows.

Firstly, there are $A$ 
four-dimensional integrations in Eq.~(\ref{eq:long}). One
integral (over $p_1$) is taken using the energy-momentum delta-function. Other $A$
integrals over energies of the spectator nucleons are taken in the
approximation that only non-relativistic nucleons with positive energy are present
in the nuclear wave function. This enables one to take the residue
over the nucleon mass pole, 
for instance,
\begin{equation}
\int \frac{d p_2^0}{(p_2)^2-m^2+i\epsilon}=-i\frac{2 \pi}{2m} \,.
\label{eq:onshell}
\end{equation}
This procedure puts all spectator nucleons on their mass
shell. Moreover, because of the energy-momentum conservation at each
nuclear vertex, the interacting nucleon is also on the mass shell
(ignoring small nuclear binding).

Secondly, since all nucleons are on the mass shell,
the Dirac spinors in the propagators of the bound nucleons can be replaced with those
of the free ones:
\begin{equation}
\hat{p}+m=\sum_{i} u^{(i)}(p) \bar {u}^{(i)}(p) \,.
\end{equation}

Thirdly,
the covariant nuclear vertex functions $\bar{\Gamma}_{A}$ and
$\Gamma_{A}$  are
expressed in terms of the corresponding non-relativistic nuclear wave
function, $\Phi_A$, for instance
\begin{equation}
\bar {u}^{(\beta)}(p_1) \bar {u}^{(i_2)}(p_2) \dots \bar {u}^{(i_A)}(p_A)
\Gamma_{A} = \Big((2 \pi)^3 2m \Big)^{\frac{A-1}{2}} (p_1^2-m^2)
\Phi_A(p_1,\beta;p_2,i_2;\dots;p_A,i_A) \,.
\label{eq:wf}
\end{equation}
Note that the nuclear wave function is normalized to the number of
nucleons $A$.
With Eqs.~(\ref{eq:onshell})-(\ref{eq:wf}) in mind, the expression for
the matrix element in Eq.~(\ref{eq:long}) simplifies
\begin{equation}
\langle P_{A}+q|J^{\mu}(0)|P_{A}\rangle= \sum_{{\rm nucleons}} \int
d^3 p_i \sum_{\alpha,\beta} \Phi_A^{\ast}(p_1+q,\alpha)
  \Phi_A(p_1,\beta) \bar {u}^{(\alpha)}(p_1+q)\hat{\Gamma}_{e.m.}^{\mu}(q)
      {u}^{(\beta)}(p_1) \,.
\label{eq:long2}
\end{equation}
Here $d^3 p_i$ denotes $A-1$ 3-dimensional integrals, $d^3 p_i=d^3
p_2 \dots d^3 p_A$. Also we  keep  only the momentum and spin indices
of the struck nucleon in the argument of the nuclear wave function;
summation over the polarization of the spectator nucleons is assumed. 

For a spinless nucleus, only the terms with $\alpha=\beta$ survive in
the product $\Phi_A^{\ast}(p_1+q,\alpha)
\Phi_A(p_1,\beta)$. Introducing the nuclear wave function averaged
over polarization of its nucleons, $\Phi_A(p_1)$, 
\begin{equation}
\Phi_A(p_1,\beta)=\frac{1}{\sqrt{2}}\Phi_A(p_1)\,,
\end{equation}  
we arrive at
\begin{eqnarray}
\langle P_{A}+q|J^{\mu}(0)|P_{A}\rangle&=&\frac{1}{2} \sum_{{\rm
    nucleons}} \int
d^3 p_i \Phi_A^{\ast}(p_1+q) \Phi_A(p_1)
\sum_{\alpha} \bar {u}^{(\alpha)}(p_1+q)\hat{\Gamma}_{e.m.}^{\mu}(q)
      {u}^{(\alpha)}(p_1) \nonumber\\
&=&\frac{1}{2} \sum_{{\rm nucleons}} \int d^3 p_i \Phi_A^{\ast}(p_1+q) \Phi_A(p_1)
     {\rm Tr}\Big(\hat{\Gamma}_{e.m.}^{\mu}(q) \hat{P}(p_1,q)\Big) \,,
\label{eq:long3}
\end{eqnarray}
where $\hat{P}(p_1,q)$ is some sort of ''off-forward propagator'', 
$\hat{P}(p_1,q)=\sum_{\alpha} {u}^{(\alpha)}(p_1) \bar {u}^{(\alpha)}(p_1+q)$. 
This propagator can be decomposed in the basis of Dirac matrices with
the result
\begin{equation}
\hat{P}(p_1,q)=\hat{p}_1+\hat{q}/2+m+\hat{p}_1\hat{q}/(2m) \,.
\label{eq:ofp}   
\end{equation}
Using the definition of $\hat{\Gamma}_{e.m.}^{\mu}$ in terms of the form
factors $F_1(t)$ and $F_2(t)$
and then taking the trace of Dirac matrices in Eq.~(\ref{eq:long3}),
we obtain
\begin{equation}
\langle P_{A}+q|J^{\mu}(0)|P_{A}\rangle=(2P_{A}+q)^{\mu}F^{e.m.}_{A}(t)=\sum_{{\rm nucleons}}\int
d^3 p_i \Phi_A^{\ast}(p_1+q) \Phi_A(p_1) (2p_1+q)^{\mu} G_E(t)
 \,,
\label{eq:long4}
\end{equation}
where the nucleon charge form factor is introduced,
$G_E(t)=F_1(t)+t/(4 m^2)F_2(t)$. As a result of the non-relativistic
reduction, Eq.~(\ref{eq:long4}) is intrinsically reference frame
dependent
and it is good only with accuracy ${\cal O}(p_1^2/m^2)$. For a more consistent 
treatment, one would need to use the light-cone
(infinite-momentum frame)  description of the
nuclear wave function, see reviews in \cite{FS81,Miller}.

In the non-relativistic for $t=0$  we can use with accuracy 
 ${\cal O}(p_1^2/m_N^2)$ the $\mu=0$ component of Eq.~(\ref{eq:long4})
to fix normalization of the wave function
\begin{equation}
F^{e.m.}_{A}(0)= \frac{m}{M_A}\sum_{{\rm nucleons}} \int d^3 p_i
|\Phi_A(p_1)|^2  G_E(0)=\frac{ZG_E^p(0)}{A} \int d^3 p_i
|\Phi_A^{\ast}(p_1)|^2 =Z \,.
\label{eq:long5}
\end{equation}
Here we introduced the proton ($G_E^p$) and neutron ($G_E^n$)
charge form factors with the properties $G_E^p(0)=1$ and $G_E^n(0)=1$.
We also used $p_1^0 \approx m$ since 
the accuracy of the non-relativistic 
convolution approximation is ${\cal O}(p_1^2/m_N^2)$.

\section{DVCS on a spinless nuclear target}

The amplitude for deeply virtual Compton scattering (DVCS) on 
any hadronic target is a function of 5 variables: $\xi$, $Q^2$, $t$, $\phi$ and
$\phi_{\perp}$. The parameter  $\xi$ is an analog of the Bjorken
variable $x_{Bj}$,
\begin{equation}
2 \xi=\frac{Q^2}{2 {\bar p} q}=\frac{2 x_{Bj}}{2-x_{Bj}+\frac{\Delta^2}{Q^2}x_{Bj}} \,,
\label{eq:xi}
\end{equation}
where ${\bar p}=(p+p^{\prime})/2=p+\Delta/2$ with $p$ and $p^{\prime}$
the momenta of the hadron in the initial and final states,
$\Delta=p^{\prime}-p$ and $\Delta^2=t$. The angle $\phi$ is the angle between the lepton
and hadron scattering planes; the angle $\phi_{\perp}$ is associated
with the target polarization and, hence, it does not appear in the
present  discussion: The amplitudes considered below are averaged
over $\phi_{\perp}$ (see Ref.~\cite{BMK02} for the detailed discussion
of kinematics of DVCS).
In the following  we shall keep an explicit dependence on
the variables $\xi$ and $t$ only.

We are interested  in sufficiently large $x_{Bj}, $ $x_{Bj} \geq 0.05$, where 
nuclear modifications of the parton densities are small and where 
average longitudinal  distances in the DVCS process are comparable or smaller
than the average internucleon distances.
In this case, the use of the impulse approximation is well justified. In this
approximation, for
a spinless nuclear target, the DVCS amplitude, $T^{\mu \nu}_A$,
 can be read off immediately from Eq.~(\ref{eq:long3})
\begin{equation}
T^{\mu \nu}_A(\xi,t)=\frac{1}{2}\sum_{{\rm nucleons}} \int d^3 p_i \Phi_A^{\ast}(p_1+\Delta) \Phi_A(p_1)
     {\rm Tr}\Big(\hat{T}^{\mu \nu}(\xi^{\prime},t) \hat{P}(p_1,q)\Big) \,,
\label{eq:d1}
\end{equation}
where $\hat{T}^{\mu \nu}$ is the amputated DVCS amplitude (without the
external spinor lines) for the free nucleon \cite{GPV01}, 
\begin{equation}
\hat{T}^{\mu \nu}(\xi^{\prime},t)=\frac{1}{2}\big(-g^{\mu \nu})_{\perp} \int^{1}_{-1}
dx C^{+}(x,\xi^{\prime}) \Big(H(x,\xi^{\prime},t)\hat{n}+E(x,\xi^{\prime},t)/(2m_N) i \sigma^{\mu \nu}
  \Delta _{\nu} \Big)+\dots \,.
 \label{eq:d2}
\end{equation}
Here, $C^{+}(x,\xi^{\prime})=1/(x-\xi^{\prime}+i\epsilon) +
1/(x+\xi^{\prime}-i\epsilon)$; $H$ and $E$ are the GPDs of the nucleon;
the ellipses denote the terms vanishing after taking the trace; $\hat{P}(p_1,q)$ is the ''off-forward'' propagator defined in
Eq.~(\ref{eq:ofp}). The variable $\xi^{\prime}$ is defined with 
respect to the interacting nucleon, 
\begin{equation}
2 \xi^{\prime}=\frac{Q^2}{(2 Ap_1+\Delta) q}=\frac{2 x_{Bj}^{\prime}
}{2-x_{Bj}^{\prime}+\frac{\Delta^2}{Q^2}x_{Bj}^{\prime}}=\frac{2 (x_{Bj}/\alpha)
}{2-(x_{Bj}/\alpha)+\frac{\Delta^2}{Q^2}(x_{Bj}/\alpha)}  \,,
\label{eq:xiprime}
\end{equation}
where we introduced the Bjorken variable defined with respect to the
struck nucleon, $x_{Bj}^{\prime}=Q^2/(2 Ap_1 q)$, and connected it
with $x_{Bj}$ by introducing the factor
$\alpha=x_{Bj}/x_{Bj}^{\prime}=A(p_1 q)/(P_{A}q)$.
Combining
Eqs.~(\ref{eq:xi}) and  (\ref{eq:xiprime}) gives
\begin{equation}
\xi^{\prime}=\frac{\xi}{\alpha} \times
\frac{2-x_{Bj}+\frac{\Delta^2}{Q^2}x_{Bj}}{2-(x_{Bj}/\alpha)+\frac{\Delta^2}{Q^2}(x_{Bj}/\alpha)} \,.
\label{eq:xixiprime}
\end{equation}
The deviation of $\alpha$ from unity and, thus, $\xi^{\prime}$ from
$\xi$, is a measure of the importance of the Fermi motion of the bound
interacting nucleon. Since $\alpha=1+{\cal O}(|p_1|/m)$, it is
legitimate to study the effects of $\alpha \neq 1$ within the impulse approximation. 

Substituting Eq.~(\ref{eq:d2}) into Eq.~(\ref{eq:d1}), taking the trace and using the on-mass-shell condition $2 (p_1 \cdot
\Delta)+\Delta^2=0$, we obtain the following result for $T^{\mu \nu}_A$
\begin{eqnarray}
T^{\mu \nu}_A(\xi,t)&=&\sum_{{\rm nucleons}} \int d^3 p_i \Phi_A^{\ast}(p_1+\Delta) \Phi_A(p_1)
\Big[\frac{1}{2}\big(-g^{\mu \nu})_{\perp}  
\int^{1}_{-1}
dx C^{+}(x,\xi^{\prime}) \nonumber\\
&\times&
\Big(H(x,\xi^{\prime},t)+E(x,\xi^{\prime},t)\frac{\Delta^2}{4m} \Big)\Big(2(p_1 \cdot
n)+(\Delta \cdot n)\Big) \Big] \,.
\label{eq:d3}
\end{eqnarray}

The light-like four-vector $n$ is completely defined by the vectors
${\bar P}_{A}$ and $q$ \cite{GPV01}
\begin{equation}
n=\frac{q+2 \xi {\bar P}_{A}}{Q^2/(4 \xi)+{\bar M}_{A}^2 \xi} \,,
\label{eq:n}
\end{equation}
with ${\bar M}_{A}^2=M_A^2-\Delta^2/4$.
Then,
\begin{equation}
p_1n=\frac{p_1q+2 \xi p_1{\bar P}_{A}}{Q^2/(4 \xi)+{\bar M}_{A}^2
 \xi}=\frac{Q^2/(2x)(\alpha/A)+2\xi P_{A}p_1-\xi \Delta^2 /2}{Q^2/(4 \xi)+{\bar M}_{A}^2
 \xi}  \,
\label{eq:p1n}
\end{equation}
and 
\begin{equation}
\Delta n=-2\xi
\label{eq:deltan} \,.
\end{equation}
Thus, the final expression for the amplitude of DVCS on a spin-0
nucleus is given by the convolution formula,
 Eq.~(\ref{eq:d3}), with $\xi^{\prime}$, $p_1 n$ and
$\Delta n$ given by Eqs.~(\ref{eq:xixiprime}), (\ref{eq:p1n}) and (\ref{eq:deltan}).

On the other hand, the DVCS amplitude  on a spinless nuclear target  can
be, at leading twist, expressed in terms of a single GPD, $H_A$, 
(the Lorentz structure of the DVCS amplitude is the same for all
spinless hadrons, and the amplitude for the pion target can be found, for
example, in \cite{BMKS01}) 
\begin{equation}
T^{\mu \nu}_A(\xi,t)=-(g^{\mu \nu})_{\perp} \int^{1}_{-1} dx C^{+}(x,\xi)
H_A(x,\xi,t) \,.
\label{eq:d5}
\end{equation}
Equating Eqs.~(\ref{eq:d3}) and (\ref{eq:d5}), we find the  
connection between nuclear and nucleon generalized parton
distributions in the form
\begin{eqnarray}
&&\int^{1}_{-1} dx C^{+}(x,\xi)
H_A(x,\xi,t)=
\frac{1}{2}\int d^3 p_i \Phi_A^{\ast}(p_1+\Delta) \Phi_A(p_1)
\int^{1}_{-1}
dx C^{+}(x,\xi^{\prime}) \nonumber\\
&&\times
\Big(Z(H^p(x,\xi^{\prime},t)+E^p(x,\xi^{\prime},t)\frac{\Delta^2}{4m})+(A-Z)(H^n(x,\xi^{\prime},t)+E^n(x,\xi^{\prime},t)\frac{\Delta^2}{4m})
\Big)
\nonumber\\
&&\times
\Big(2(p_1 \cdot
n)+(\Delta \cdot n)\Big) \,,
\label{eq:d6}
\end{eqnarray}
where the superscripts ''$p$'' and ''$n$'' denote the proton and neutron GPDs.

Equation~(\ref{eq:d6}) has both real and imaginary parts. They can be
separated with the result
\begin{eqnarray}
&&\int^{1}_{-1} dx \Big(P(\frac{1}{x-\xi})+P(\frac{1}{x+\xi})\Big)
H_A(x,\xi,t)=
\frac{1}{2}\int d^3 p_i \Phi_A^{\ast}(p_1+\Delta) \Phi_A(p_1) \nonumber\\
&&\times \int^{1}_{-1} dx \Big(P(\frac{1}{x-\xi^{\prime)}})+P(\frac{1}{x+\xi^{\prime)}})\Big) 
\Big(Z(H^p(x,\xi^{\prime},t)+E^p(x,\xi^{\prime},t)\frac{\Delta^2}{4m}) \nonumber\\
&&+(A-Z)(H^n(x,\xi^{\prime},t)+E^n(x,\xi^{\prime},t)\frac{\Delta^2}{4m})\Big) \Big(2(p_1 \cdot
n)+(\Delta \cdot n)\Big) \,
\label{eq:d7}
\end{eqnarray}
and 
\begin{eqnarray}
&&H_A(\xi,\xi,t)-H_A(-\xi,\xi,t)=
\frac{1}{2}\int d^3 p_i \Phi_A^{\ast}(p_1+\Delta) \Phi_A(p_1) \nonumber\\
&&\times
\Big(Z(H^p(\xi^{\prime},\xi^{\prime},t)-H^p(-\xi^{\prime},\xi^{\prime},t)+(E^p(\xi^{\prime},\xi^{\prime},t)-E^p(-\xi^{\prime},\xi^{\prime},t))\frac{\Delta^2}{4m})
\nonumber\\
&&+(A-Z)(H^n(\xi^{\prime},\xi^{\prime},t)-H^n(-\xi^{\prime},\xi^{\prime},t)+(E^n(\xi^{\prime},\xi^{\prime},t)-E^n(-\xi^{\prime},\xi^{\prime},t))\frac{\Delta^2}{4m})
\Big) \nonumber\\
&&\times \Big(2(p_1 \cdot n)+(\Delta \cdot n)\Big) \,,
\label{eq:d8}
\end{eqnarray}
where $P(1/x)$ denotes the principal value integral.

\section{Beam-charge and single-spin asymmetries}

Generalized parton distribution functions are great in number and
depend on many variables. Thus, one studies special observables
(asymmetries) involving GPDs aiming to study only certain aspects of
GPDs and only a few at a time. In this section we consider the beam-charge and
single-spin asymmetries for DVCS on a spinless nuclear target.

The beam-charge asymmetry, $A_{C}$, is measured by scattering
unpolarized leptons of opposite charges (positrons and electrons) on
unpolarized hadronic targets.
The asymmetry is then defined as the function of the angle $\phi$, the
angle between the leptonic and hadronic scattering planes (the
dependence on $Q^2$, $\xi$ and $t$ is implicit),
\begin{equation}
A_{C}(\phi)=\frac{N^{+}(\phi)-N^{-}(\phi)}{N^{+}(\phi)+N^{-}(\phi)}\,,
\label{eq:ac1}
\end{equation}
where $N^{+}$ ($N^{-}$) is proportional to the scattering cross
section for the incoming positron (electron).
It is important to note that  the DVCS amplitude, ${\cal T}_{DVCS}$, interferes with
the purely electromagnetic  Bethe-Heitler amplitude, ${\cal T}_{BH}$, so that the DVCS
asymmetries depend and, in some kinematics, are dominated by
interference of both amplitudes, ${\cal I}$. For instance, 
the asymmetry $A_{C}$ probes the real part of the interference between
the DVCS and Bethe-Heitler amplitudes.

The expressions for the DVCS, Bethe-Heitler and interference
contributions to the scattering cross section on the pion target
(since the
pion can be replaced by any spin-0 hadronic target, the expressions
for the asymmetries are also valid for spinless nuclei)
 are derived in \cite{BMKS01}. The beam-charge asymmetry can be
 written as
\begin{equation}
A_{C}(\phi)=\frac{{\cal I}(\lambda=1)+{\cal I}(\lambda=-1)}{2|{\cal
    T}_{BH}|^2+{\cal I}(\lambda=1)+{\cal I}(\lambda=-1)+{\cal
    T}_{DVCS}(\lambda=1)+{\cal T}_{DVCS}(\lambda=-1)} \,,
\label{eq:ac2}
\end{equation}
where $\lambda$ the helicity of the incoming (massless) lepton and
\begin{eqnarray}
&&{\cal I}(\lambda)=-\frac{F^{e.m.}(t)}{x_{Bj}^2 y^3  \Delta
 ^2 {\cal P}_1 {\cal P}_2} \Big(\frac{\Delta^2}{Q^2} c_0^{{\cal I}}+
 \sum_{m=1}^{2} K^m (c_m^{{\cal I}}\cos(m\phi)+\lambda s_m^{{\cal I}}
 \sin(m\phi))+\frac{Q^2}{M^2}K^3 c_3^{{\cal I}} \cos(3 \phi) \Big) \,, \nonumber\\
&&|{\cal T}_{BH}|^2=-\frac{(F^{e.m.}(t))^2}{x_{Bj}^2 y^2 (1+\epsilon^2) \Delta
 ^2 {\cal P}_1 {\cal P}_2} \sum_{m=0}^{2} c_m^{BH} K^m \cos(m\phi ) \,,\nonumber\\
&&|{\cal T}_{DVCS}(\lambda)|^2=\frac{1}{x_{Bj} y^2 Q^2}
 \Big(c_0^{DVCS}+K(c_1^{DVCS} \cos(\phi)+\lambda
 s_1^{DVCS}\sin(\phi)) \nonumber\\
&&+ \frac{Q^2}{M^2}K^2 c_2^{DVCS} \cos(2 \phi) \Big)
 \,.
\label{eq:ac3}
\end{eqnarray}
Here ${\cal P}_1$ and ${\cal P}_2$ are dimensionless lepton propagators (divided
  by $Q^2$); $K$ and $\epsilon^2$ and kinematics factors; the
  coefficients $c_{i}$ and $s_{i}$ are given by Eqs.~(31), (32), (33)
  and (11) of Ref.~\cite{BMKS01}. Note that while the result of
  Ref.~\cite{BMKS01} includes also the twist-three terms, we consider
  only the leading, twist-two, contribution.

In general,  exact Eqs.~(\ref{eq:d7}) and (\ref{eq:d8}) enable one to
evaluate $A_{C}$ in the most general case.
However, the main goal of this paper is to write down simple and yet
reliable expressions, where the effects associated with the nuclear
target are presented in a transparent form. To this end, let us
consider the following kinematics for DVCS on a nuclear target:
$\Delta^2=t \ll Q^2$, $Q^2$ equals a few GeV$^2$ and $x_{Bj} \geq 0.1$. 
These conditions correspond to the kinematics of the HERMES experiment at DESY. 
In this kinematics, the Bethe-Heitler process dominates the cross
section and, moreover, we can neglect the terms proportional 
$\epsilon^2$ and $\Delta^2/Q^2$ and keep only the leading terms in
powers of $K$ ($K \propto \sqrt{\Delta}$). The simplified expression for the beam-charge asymmetry
reads
\begin{eqnarray}
A_{C}(\phi)&=&\cos(\phi) \frac{K c_1^{{\cal I}}}{y c_0^{BH} F_{A}^{e.m.}(t)} \nonumber\\
&=&-\cos(\phi) \frac{K 8(2-2y+y^2)x_{Bj}}{y c_0^{BH}} \frac{\int^{1}_{-1} dx \Big(P(\frac{1}{x-\xi})+P(\frac{1}{x+\xi})\Big)
H_A(x,\xi,t)}{F_{A}^{e.m.}(t)} \,,
\label{eq:ac4}
\end{eqnarray}
where $\int^{1}_{-1} dx \Big(P(\frac{1}{x-\xi})+P(\frac{1}{x+\xi})\Big)H_A(x,\xi,t)$ is given by Eq.~(\ref{eq:d7}) and the nuclear charge
form factor $F_{A}^{e.m.}(t)$ can be parametrized phenomenologically
using the experimental data on the nuclear charge radius
  or, alternatively, can be evaluated using Eq.~(\ref{eq:long4}).
In order to get a first estimate of the influence of nuclear medium on
$A_{C}$, in the limit $t=0$ let us evaluate the asymmetry ignoring the 
Fermi motion effects 
so that $\alpha=1$.
Neglecting the ${\cal O}(x_{Bj})$ effects in the definition of $\xi$ in Eq.~(\ref{eq:xi}) and
the ${\cal O}(\xi^2)$ effects in the definition of $p_1 n$ in
Eq.~(\ref{eq:p1n}), one has $p_1 n=\alpha/A=1/A$. Then the nominator
of the last term in Eq.~(\ref{eq:ac4}) (see Eq.~(\ref{eq:d7})) becomes
\begin{eqnarray}
&&\int^{1}_{-1} dx
\Big(P(\frac{1}{x-\xi})+P(\frac{1}{x+\xi})\Big)H_A(x,\xi,0)
\nonumber\\
&&=\int^{1}_{-1}
dx
\Big(P(\frac{1}{x-\xi})+P(\frac{1}{x+\xi})\Big)\Big(ZH^p(x,\xi,0)+(A-Z)H^n(x,\xi,0)\Big)
\,,
\label{eq:ac5}
\end{eqnarray}
where in the last step we have used that $\xi^{\prime}= \xi$ in the
considered approximation.

The denominator of the last term in  Eq.~(\ref{eq:ac4}) is
$F_{A}^{e.m.}(0)=Z$. In order to quantify the resulting nuclear effects, one can
introduce the ratio of the beam-charge asymmetries measured with the
nuclear and the proton targets, $A_{C}/A_{C}^{{\rm proton}}$. Using
Eqs.~(\ref{eq:ac4}) and (\ref{eq:ac5}), the ratio can be presented in
the following form
\begin{equation}
\frac{A_{C}(\phi)}{A_{C}^{{\rm
      proton}}(\phi)}=\frac{\int^{1}_{-1} dx \Big(
      P(\frac{1}{x-\xi})+P(\frac{1}{x+\xi})\Big)\Big(H^p(x,\xi,0)+(A/Z-1)H^n(x,\xi,0)\Big)}{\int^{1}_{-1}
      dx \Big(P(\frac{1}{x-\xi})+P(\frac{1}{x+\xi})\Big)H^p(x,\xi,0)}
      \,.
\label{eq:ac6}
\end{equation}
The immediate consequence of
Eq.~(\ref{eq:ac6}) is that the ratio $A_{C}/A_{C}^{{\rm proton}}$ is
greater than unity, i.e. the beam-charge asymmetry for the nuclear
target is larger than the corresponding asymmetry for the proton.

The second kind of asymmetry, the single spin asymmetry, $A_{LU}$, is
 measured by scattering  
longitudinally polarized leptons of opposite helicities on an
 unpolarized hadronic target. The resulting asymmetry is defined as
\begin{equation}
A_{LU}(\phi)=\frac{N^{\lambda=1}(\phi)-N^{\lambda=-1}(\phi)}{N^{\lambda=1}(\phi)+N^{\lambda=-1}(\phi)}\,,
\label{eq:lu1}
\end{equation}
where $N^{\lambda=1}$ ($N^{\lambda=-1}$) is proportional to the scattering cross
section for the incoming positron with positive (negative
helicity). This asymmetry probes the
 imaginary part of the interference between the DVCS and Bethe-Heitler
 amplitudes. In the notations of Eq.~(\ref{eq:ac2}), $A_{LU}$
 can be written in the  form
\begin{equation}
A_{LU}(\phi)=\frac{{\cal I}(\lambda=1)-{\cal I}(\lambda=-1)}{2|{\cal
    T}_{BH}|^2+{\cal I}(\lambda=1)-{\cal I}(\lambda=-1)+{\cal
    T}_{DVCS}(\lambda=1)-{\cal T}_{DVCS}(\lambda=-1)} \,.
\label{eq:lu2}
\end{equation}

In the approximation that the Bethe-Heitler process dominates the
cross section, a simplified expression for the single-spin asymmetry
can be presented
\begin{eqnarray}
A_{LU}(\phi)&=&\sin(\phi) \frac{K s_1^{{\cal I}}}{y c_0^{BH} F_{A}^{e.m.}(t)} \nonumber\\
&=&\sin(\phi) \frac{K 8y(2-y)x_{Bj}}{y c_0^{BH}} \frac{H_A(\xi,\xi,t)-H_A(-\xi,\xi,t)}{F_{A}^{e.m.}(t)} \,,
\label{eq:lu3}
\end{eqnarray}
where $H_A(\xi,\xi,t)-H_A(-\xi,\xi,t)$ is given by Eq.~(\ref{eq:d8}).
In order to obtain a rough estimate of the influence of nuclear
effects on $A_{LU}$, the latter can be evaluated in the limit $t=0$
and $\xi^{\prime}=\xi$ using Eq.~(\ref{eq:d8}). Then one immediately
obtains for the ratio of the nuclear to the proton asymmetries,
$A_{LU}/A_{LU}^{{\rm proton}}$,
\begin{equation}
\frac{A_{LU}(\phi)}{A_{LU}^{{\rm
      proton}}\phi)}=\frac{H^p(\xi,\xi,0)-H^p(-\xi,\xi,0)+(A/Z-1)\Big(H^n(\xi,\xi,0)-H^n(-\xi,\xi,0)\Big)}{H^p(\xi,\xi,0)-H^p(-\xi,\xi,0)} \,.
\label{eq:lu4}
\end{equation}  
Again, like in the case of $A_{C}/A_{C}^{{\rm proton}}$, one concludes
that the single-spin asymmetry for the nuclear target is larger than
that for the proton.

Similarly to the case of the matrix element of the electromagnetic
current considered in Sec.~\ref{sec:em}, Eqs.~(\ref{eq:d7}) and
(\ref{eq:d8})
are reference frame dependent: The three-vector $\Delta$ entering the
argument of the nuclear wave function depends on the reference frame.
This is an intrinsic problem of the impulse approximation.  
In what follows we shall use the laboratory reference frame and choose
$q=(q^0,0,0,-|q^z|)$. Then the variable $\alpha=1+p_1^z/m$ and
from the condition $(q-\Delta)^2=0$, we
obtain that $\Delta^z \approx -x_{bj} m$.

Having fixed the kinematics, we can  discuss two important effects
not included in the approximate expressions of Eqs.~(\ref{eq:ac4}) and 
(\ref{eq:lu3}).
Firstly, our exact impulse approximation expressions,
Eqs.~(\ref{eq:d7}) and (\ref{eq:d8}), describe non-static bound
nucleons so that, on average, $\alpha >1$ (since $\Delta^z< 0 $, the
nuclear wave function favors positive $p_{1}^z$),
and, hence, 
$\xi^{\prime} < \xi$. This means that
the GPDs of the bound nucleons, $H$ and $E$, are probed at smaller
values of variable $\xi$ which leads to an additional enhancement of
the ratios $A_{C}/A_{C}^{{\rm proton}}$ and $A_{LU}/A_{LU}^{{\rm
    proton}}$ on the top of the ''combinatoric'' enhancement by the
term $(A/Z-1)H^n$. 

Secondly, our final expressions for the
asymmetries, Eqs.~(\ref{eq:ac4}) and  (\ref{eq:lu3}),
which are obtained in the
approximation of a small momentum
transfer $\Delta$ and dominance of the Bethe-Heitler process, 
assume that
DVCS on a nuclear target is coherent, i.e. the target remains
intact. 
However, as soon as $t \neq 0$, both coherent and incoherent
(nucleus breaks up) contributions enter the total cross section so
that the expressions for $A_{C}$ and $A_{LU}$ should be modified
accordingly. In order to achieve this, we generalize the expression
for the sum of the incohent and coherent contributions to the cross
section of  pion-nucleus production of two jets \cite{FSM} to the case
of DVCS on nuclei
\footnote{The main difference from Ref.~\cite{FSM} is that in the
  present case
one deals with interference of two amplitudes: One is coupled to the protons only
and another is coupled to all nucleons with approximately the same
  strength.}.
\vskip -1cm
\begin{figure}[h]
\begin{center}
\epsfig{file=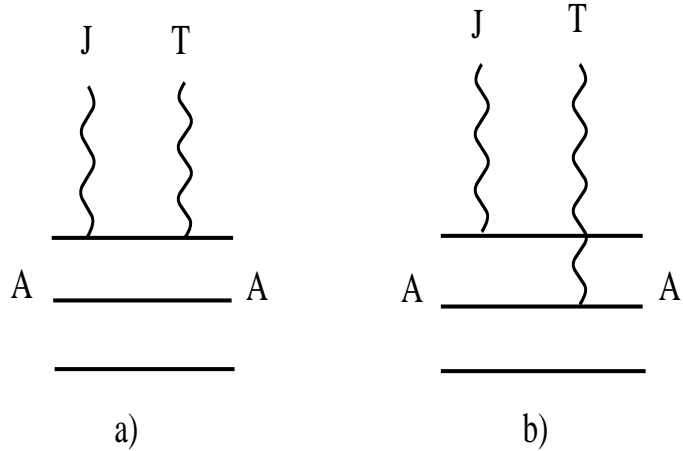,width=13cm,height=13cm}
\vskip -5cm
\caption{The schematic representation of the interference between the
 Bethe-Heitler ($J$) and DVCS ($T$) amplitudes on nuclei: There are
 $Z$ attachments of both $J$ and $T$ to the same proton
 (a) and $Z(A-1)$ attachments of $J$ to the proton and $T$ to a
 different nucleon (b).}
\end{center}
\end{figure}

 The modified asymmetries become
\begin{eqnarray}
&&A_{C}(\phi)=-\cos(\phi) \frac{K 8(2-2y+y^2)x_{Bj}}{y c_0^{BH}}
  \nonumber\\
&&\times
 \frac{\int^{1}_{-1} dx \Big(P(\frac{1}{x-\xi})+P(\frac{1}{x+\xi})\Big)\Big(
Z H^p(x,\xi,t)+Z(A-1)F_{A}^{e.m.}(t^{\prime}) H_A(x,\xi,t^{\prime})\Big)}{\Big(ZF_1(t)+Z(Z-1)(F_{A}^{e.m.}(t^{\prime}))^2\Big)} \,
\label{eq:acfull}
\end{eqnarray}
and
\begin{eqnarray}
&&A_{LU}(\phi)=\sin(\phi) \frac{K 8y(2-y)x_{Bj}}{y c_0^{BH}}
  \nonumber\\
&&\times
 \frac{\Big(
Z (H^p(\xi,\xi,t)-H^p(-\xi,\xi,t))
+Z(A-1)F_{A}^{e.m.}(t^{\prime}) (H_A(\xi,\xi,t^{\prime})-H_A(-\xi,\xi,t^{\prime}))
\Big)}{\Big(ZF_1(t)+Z(Z-1)(F_{A}^{e.m.}(t^{\prime}))^2\Big)} \,
\label{eq:lufull}
\end{eqnarray}
where $t^{\prime}=tA/(A-1)$. The schematic representation of the
origin of the combinatoric factors $Z$ and $Z(A-1)$ is given in Fig.~1.
The first terms in the nominator and
denominator of Eqs.~(\ref{eq:acfull}) and  (\ref{eq:lufull}) describe
the 
 contribution coming from the attachment in the ``in'' and ``out'' states
to the same nucleon (it gives the dominant incoherent term at large
$t$) that,
  at small $t$ (neglecting the
neutron contribution suppressed by the smallness of the
electromagnetic form factors and also neglecting the contribution of
 the GPDs $E$),
 is proportional to the number
of protons, $Z$, times the GPD $H$ of the free proton. This contribution
has a slow $t$-dependence governed by the proton elastic form factor
$F_1(t)$. The 
contribution given by the second term in the
nominator 
and denominator of Eqs.~(\ref{eq:acfull}) and
(\ref{eq:lufull}) is due to the attachment to two different nucleons.
It is mostly coherent and it 
 has a much steeper $t$-dependence dictated essentially
by the nuclear charge form factor $F_{A}^{e.m.}(t)$.

The main point of considering Eqs.~(\ref{eq:acfull}) and
(\ref{eq:lufull}) is the following. If the experimental equipment does
not allow to extract the purely coherent DVCS
so that Eqs.~(\ref{eq:ac4}) and (\ref{eq:lu3}) can be used,
the experimental asymmetries
present a sum of the coherent and incoherent
contributions and are given by Eqs.~(\ref{eq:acfull}) and (\ref{eq:lufull}). While  $A_{C}/A_{C}^{{\rm proton}}$ and
$A_{LU}/A_{LU}^{{\rm proton}}$  are significantly larger than unity for
 coherent nuclear DVCS (the ratios of the asymmetries are close to
the factor of 
two in the considered kinematics),
$A_{C}/A_{C}^{{\rm proton}}=A_{LU}/A_{LU}^{{\rm proton}}=1$ for the
 incoherent part. Thus, the inclusion of the incoherent contribution
 decreases the ratio of the asymmetries.

\begin{figure}[t]
\begin{center}
\epsfig{file=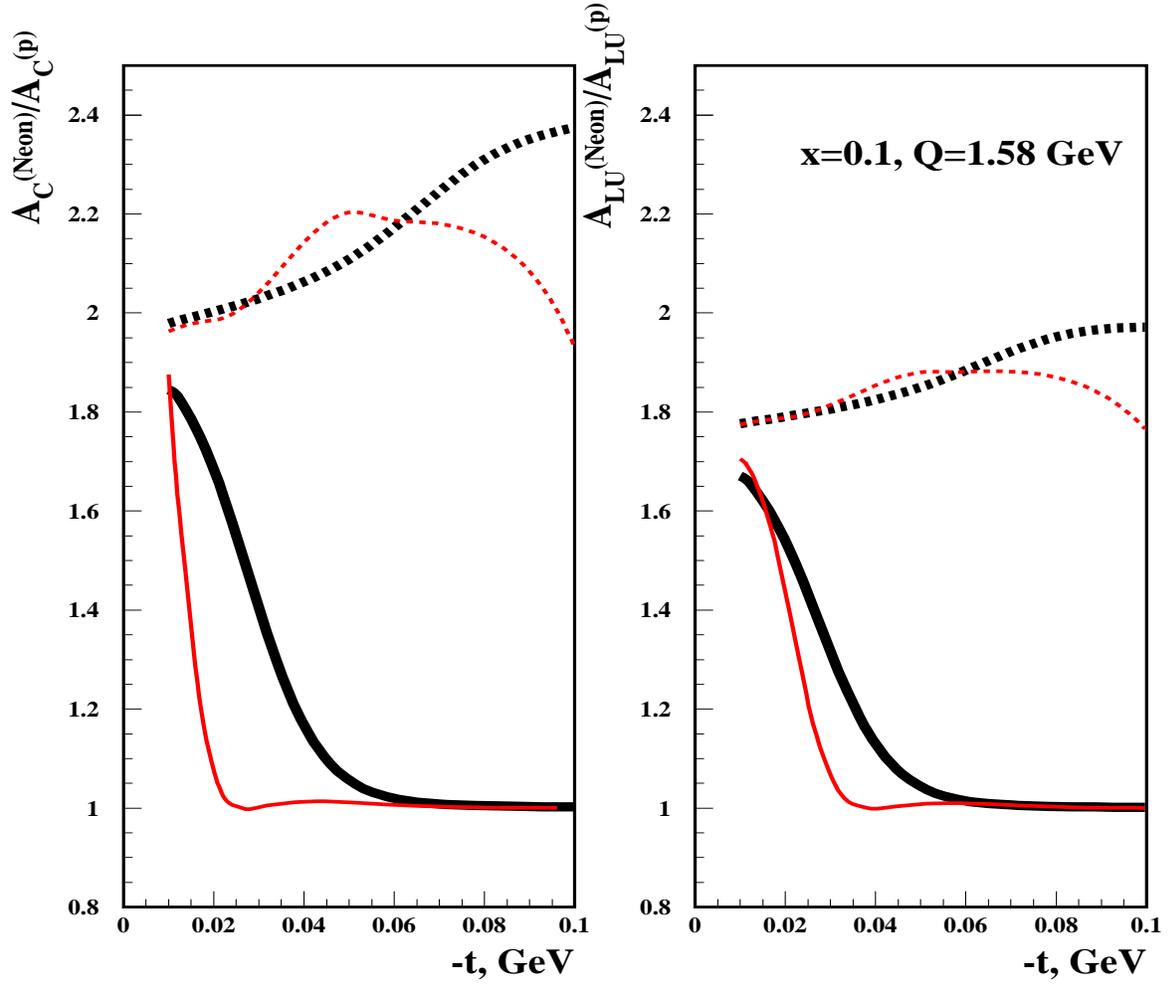,width=16cm,height=14cm}
\caption{The ratio of nuclear to proton asymmetries $A_{C}/A_{C}^{{\rm
      proton}}$ and $A_{LU}/A_{LU}^{{\rm proton}}$ for Neon (thick black)
      and Krypton (thin red).}
\end{center}
\end{figure}

In order to illustrate these 
points,
 we consider DVCS on  spinless
nuclei of Neon ($A=20$ and $Z=10$) and Krypton (spinless isotope with $A=76$ and $Z=36$).
Figure~2 presents the ratio of
the nuclear to proton asymmetries, $A_{C}/A_{C}^{{\rm proton}}$ and
$A_{LU}/A_{LU}^{{\rm proton}}$, as a function of $t$ at $x=0.1$ and
$Q^2=2.5$ GeV$^2$ and fixed $\cos(\phi)$. This choice of $x$ and $Q^2$
roughly corresponds to the kinematics of the HERMES DVCS experiment with
nuclei
\cite{Ellinghaus}. The solid curves is the full result
including both incoherent and coherent contributions; the dashed
curves include only the coherent part of the cross section. 
The following two features of Fig.~2 are of interest. Firstly, the
ratio of the asymmetries for the coherent contribution (dashed curves)
is significantly larger than unity. This is expected from the
approximate expressions of Eqs.~(\ref{eq:ac6}) and (\ref{eq:lu4}). As
discussed above, an
additional $t$-dependent enhancement arises because the bound nucleon
GPDs, which enter the complete expressions of Eqs.~(\ref{eq:d7}) and
(\ref{eq:d8}), are probed at smaller values of $\xi$ than for the free
proton.

Secondly, the inclusion of the incoherent contribution significantly
reduces the ratio of the asymmetries (solid curves), especially at
larger values of $|t|$, where the coherent contribution is suppressed
by the smallness of the nuclear form factor. From the practical point of view,
this means that if the experimental resolution in $t$ is poor, the
asymmetries extracted from the experiment are obtained 
from the $t$-averaged data.
This means for Eqs.~(\ref{eq:acfull})
and (\ref{eq:lufull}) that first one integrates separately
the nominator and
denominator  over $t$  and only then the
ratio is taken.
Integrating the nominators and denominators of $A_C$, $A_{LU}$, $A_{C}^{{\rm proton}}$ and
$A_{LU}^{{\rm proton}}$ over $|t|$ from $|t|_{{\rm min}}=x_{Bj}
m_N/(1-x_{Bj}+x_{Bj}m_N^2/Q^2)$ to $|t|_{{\rm max}}=0.1$ GeV for Neon (the maximal $|t|$
measured by HERMES \cite{Ellinghaus}) and to $|t|_{{\rm max}}=0.05$
GeV for Krypton (the coherent contribution for Krypton occupies a
smaller $t$-range),
and then taking the
ratio of the nuclear to proton asymmetries, we obtain for Neon
\begin{eqnarray}
&&\frac{\langle A_C(\phi,t) \rangle }{\langle A_C^{{\rm
        proton}}(\phi,t) \rangle}=1.10 \pm 0.01 \,,\nonumber\\
&&\frac{\langle A_{LU}(\phi,t) \rangle}{\langle A_{LU}^{{\rm
        proton}}(\phi,t)\rangle}=1.05 \pm 0.01 \,,
\label{eq:res1}
\end{eqnarray}
and for Krypton
\begin{eqnarray}
&&\frac{\langle A_C(\phi,t) \rangle }{\langle A_C^{{\rm
        proton}}(\phi,t) \rangle}=1.27 \pm 0.01 \,,\nonumber\\
&&\frac{\langle A_{LU}(\phi,t) \rangle}{\langle A_{LU}^{{\rm
        proton}}(\phi,t)\rangle}=1.18 \pm 0.01 \,,
\label{eq:res2}
\end{eqnarray}
where
\begin{eqnarray}
&&\langle A_C(\phi,t) \rangle=-\cos(\phi) \frac{8(2-2y+y^2)x_{Bj}}{y c_0^{BH}}
  \nonumber\\
&&\times
 \frac{\int^{|t|_{{\rm max}}}_{|t|_{{\rm min}}} d|t| K \int^{1}_{-1} dx \Big(P(\frac{1}{x-\xi})+P(\frac{1}{x+\xi})\Big)\Big(
Z H^p(x,\xi,t)+Z(A-1)F_{A}^{e.m.}(t^{\prime})
  H_A(x,\xi,t^{\prime})\Big)}{\int^{|t|_{{\rm max}}}_{|t|_{{\rm min}}}
  d|t|\Big(ZF_1(t)+Z(Z-1)(F_{A}^{e.m.}(t^{\prime}))^2\Big)} \,,\nonumber\\
&&\langle A_C^{{\rm proton}}(\phi,t) \rangle=-\cos(\phi) \frac{8(2-2y+y^2)x_{Bj}}{y c_0^{BH}} \nonumber\\
&&\times
 \frac{\int^{|t|_{{\rm max}}}_{|t|_{{\rm min}}} d|t| K \int^{1}_{-1} dx \Big(P(\frac{1}{x-\xi})+P(\frac{1}{x+\xi})\Big) H^p(x,\xi,t)}{\int^{0.1}_{|t|_{{\rm min}}}
  d|t|ZF_1(t)} \,.
\label{eq:def1}
\end{eqnarray}
The quantity $\langle A_{LU}(\phi,t) \rangle$ is defined similarly to
$\langle A_{C}(\phi,t) \rangle$ with evident substitutions.

In order to carry out the above numerical analysis, we used a double
distribution parametrization for the GPDs $H^u$ and $H^d$ without the D-term
\begin{equation}
H^q(x,\xi)=\int^{1}_{-1} d\beta \int^{1-|\beta|}_{-1+|\beta|}
d\alpha \delta(x-\beta-\alpha \xi) h(\beta,\alpha) q(\beta) \,,
\label{eq:dd}
\end{equation}
with $h(\beta,\alpha)=0.75((1-|\beta|)^2-\alpha^2)/(1-|\beta|)^3$
\cite{GPV01} and
$q(\beta)$ the parton density of the $u$ or $d$ quark. The
parametrization for $q(\beta)$ is taken as that given  by the CTEQ5M
fit \cite{CTEQ5}. The $t$-dependence is chosen as in Ref.~\cite{GPV01}:
$H^u(x,\xi,t)=H^u(x,\xi)F_1(t)$, $H^d(x,\xi,t)=H^d(x,\xi)F_1(t)$ and
$H^s(x,\xi,t)=0$, where $F_1(t)$ is the elastic form factor of the
proton. This form factor is parametrized in a dipole form
\begin{equation}
F_1(t)=\frac{1}{(1+|t|/(0.71 \, {\rm GeV}^2)^2} \,.
\label{eq:f1}
\end{equation}
The $t$-dependence of the nuclear GPD $H_A(x,\xi,t)$ is given by
Eq.~(\ref{eq:d6}). The theoretical error included in Eqs.~(\ref{eq:res1})
and (\ref{eq:res2}) reflects the uncertainty in the $t$-factorization
ansatz for $H^q(x,\xi,t)$ and the uncertainty in the slope of the
$t$-dependence of the elementary $\gamma^{\ast} N$ amplitude. The
error was assessed by the modification $F_1(t) \to  F_1(t)\exp(-2t)$:
The answer for the ratios in Eqs.~(\ref{eq:res1})
and (\ref{eq:res2}) changes very insignificantly.

The nuclear form factor is obtained
 as a Fourier transform of the nuclear density 
in the coordinate space
\begin{equation}
\rho_A(r)=\frac{\rho_0}{1+\exp((r-c)/z)}
\,,
\label{eq:wf2}
\end{equation}
where $\rho_0=0.0081124$ fm$^{-3}$, $c=2.740$ fm and $z=0.572$ fm for
Neon; $\rho_0=0.0020925$ fm$^{-3}$, $c=4.649$ fm and $z=0.545$ fm for
Krypton \cite{Vries}.

\section{Conclusions}

The nuclear effect of Fermi motion in DVCS on spinless nuclear targets
is considered within the framework of the impulse approximation. The
amplitude of nuclear DVCS is expressed in terms of the convolution of
the GPDs of the nucleons with the non-relativistic nuclear wave
function.

The expressions for the beam-charge and single spin asymmetries in the
HERMES kinematics are discussed extensively. It is shown that, apart
from the combinatoric enhancement of the asymmetries because of the
neutron contribution (see Eqs.~(\ref{eq:ac4}) and (\ref{eq:lu3}),
there are two additional effects: while Fermi motion of the nucleons enhances
the asymmetries, the presence of the incoherent scattering
at $t \neq 0$ drastically reduces the asymmetries.

\section*{Acknowledgements}

We would like to thank M. Amarian, L. Frankfurt, P.V. Pobylitsa, R. Shanidze and
especially M. Polyakov for many valuable discussions and comments. The
work is supported by the Sofia Kovalevskaya Program of the Alexander
von Humboldt Foundation (Germany) and DOE (USA).


\begin{thebibliography}{0}

\bibitem{HERMES} HERMES Collab., A. Airapetian  {\it at al.},
  Phys. Rev. Lett. {\bf 87} (2001) 182001.

\bibitem{ZEUS} ZEUS Collab., {\it Measurement of the deeply virtual
  Compton scattering cross section at HERA}, abstract 564,
  Intern. Europhys. Conf. on High Energy Physics, Budapest, Hungary,
  July 12.18, 2001.

\bibitem{H1} H1 Collab., C. Adloff {\it at al.}, Phys. Lett. {\bf B
  517} (2001) 47.

\bibitem{CLAS} CLAS Collab., S. Stepanyan {\it at al.},
  Phys. Rev. Lett. {\bf 87} (2001) 182002.

\bibitem{BMK02} A.V. Belitsky, D. M{\"u}ller and A. Kirchner, {\it
  Theory of deeply virtual Compton scattering on the nucleon},
  preprint hep-ph/0112108, v2.

\bibitem{GPV01} K. Goeke, M.V. Polyakov and M. Vanderhaeghen, Prog.
  Part. Nucl. Phys. {\bf 47} (2001) 401.

\bibitem{GV98} P.A.M. Guichon and M. Vanderhaeghen, Prog.
  Part. Nucl. Phys. {\bf 41} (1998) 125.

\bibitem{Ellinghaus} F. Ellinghaus, R. Shanidze and J. Volmer, 
 {\it Deeply-Virtual Compton
  Scattering on Deuterium and Neon at HERMES}, preprint hep-ex/0212019.

\bibitem{Polyakov} M.V. Polyakov, {\it Generalized parton
  distributions and strong forces inside nucleons and nuclei},
  preprint hep-ph/0210165.

\bibitem{Cano} F. Cano and
 B. Pire, Nucl. Phys. {\bf A711} (2002) 133;
  F. Cano and B. Pire, {\it Deeply Virtual Compton Scattering on
  Spin-1 Nuclei}, preprint hep-ph/0211444.

\bibitem{FS81} L.L. Frankfurt and M.I. Strikman, Phys. Rep.{\bf 76}
(1981) 215.
 
\bibitem{Sargsian} M. Sargsian, Int. J. Mod. Phys. {\bf E10} (2001) 405. 


\bibitem{Miller}
G.A. Miller,
Prog. Part. Nucl. Phys. {\bf 45} (2000) 83.

\bibitem{BMKS01} A.V. Belitsky, D. M{\"u}ller, A. Kirchner and
  A. Sch{\"a}fer, Phys. Rev. {\bf D 64} (2001) 116002.

\bibitem{FSM} L. Frankfurt, G.A. Miller and M. Strikman,
  Phys. Rev. {\bf D 65} (2002) 094015.

\bibitem{CTEQ5} H. Lai {\it at al.}, Eur. Phys. J. {\bf C 12} (2000) 375.

\bibitem{Vries} H. De Vries, C.W. De Jager and C. De Vries, At. Data
  Nucl. Data Tables, {\bf 36} (1987) 495.  

\end{thebibliography}
\end{document}